\documentclass[conference]{IEEEtran}
\IEEEoverridecommandlockouts

\usepackage{cite}
\usepackage{amsmath,amssymb,amsfonts}
\usepackage{algorithm, algorithmic}
\usepackage{graphicx}
\usepackage{textcomp}
\usepackage{xcolor}
\usepackage{comment}
\def\BibTeX{{\rm B\kern-.05em{\sc i\kern-.025em b}\kern-.08em
    T\kern-.1667em\lower.7ex\hbox{E}\kern-.125emX}}
\begin{document}

\title{Vertical Federated Learning for Failure-Cause Identification in Disaggregated Microwave Networks}

\author{Fatih Temiz$^{1*}$, Memedhe Ibrahimi$^{1}$, Francesco Musumeci$^{1}$, Claudio Passera$^{2}$ and Massimo Tornatore$^{1}$}
\author{\IEEEauthorblockN{Fatih Temiz$^{1*}$, Memedhe Ibrahimi$^{1}$, Francesco Musumeci$^{1}$, Claudio Passera$^{2}$, and Massimo Tornatore$^{1}$}
\IEEEauthorblockA{\textit{$^1$Department of Electronics, Information and Bioengineering, Politecnico di Milano, Milan, Italy} \\ \textit{$^2$SIAE-Microelettronica, Cologno Monzese, Italy};  $^*$Corresponding author: \emph{fatih.temiz@mail.polimi.it}}
}

\maketitle

\begin{abstract}
Machine Learning (ML) has proven to be a promising solution to provide novel scalable and efficient fault management solutions in modern 5G-and-beyond communication networks. In the context of microwave networks, ML-based solutions have received significant attention. However, current solutions can only be applied to monolithic scenarios in which a single entity (e.g., an operator) manages the entire network. As current network architectures move towards disaggregated communication platforms in which multiple operators and vendors collaborate to achieve cost-efficient and reliable network management, new ML-based approaches for fault management must tackle the challenges of sharing business-critical information due to potential conflicts of interest. In this study, we explore the application of Federated Learning in disaggregated microwave networks for failure-cause identification using a real microwave hardware failure dataset. In particular, we investigate the application of two Vertical Federated Learning (VFL), namely using Split Neural Networks (\emph{SplitNNs}) and Federated Learning based on Gradient Boosting Decision Trees (\emph{FedTree}), on different multi-vendor deployment scenarios, and we compare them to a centralized scenario where data is managed by a single entity. Our experimental results show that VFL-based scenarios can achieve F1-Scores consistently within at most a 1\% gap with respect to a centralized scenario, regardless of the deployment strategies or model types, while also ensuring minimal leakage of sensitive-data.
\end{abstract}


\section{Introduction}
Microwave networks are an integral part of today's and future mobile transport networks, offering cost-efficient and high-capacity transmission to meet the rising demand for backhaul capacity \cite{Ericsson2022Microwave}. In microwave networks, effective and efficient failure management is essential. Previously, failure management was manually carried out by network experts using monitoring data collected by the Network Management System (NMS). This human-based approach results in a time-consuming and error-prone Root Cause Analysis(RCA) that can cause, e.g., costly unnecessary site visits. Machine Learning (ML)-based network failure management in microwave networks has hence gained traction in the last years \cite{Ericsson2023Microwave,10108226}, as it promises to ensure cost-efficient and accurate failure management\cite{ASAP}.

However, current solutions can only be applied to monolithic scenarios, with tightly integrated components managed by a single vendor, while next-generation \emph{5G-and-beyond} networks are moving towards disaggregated architectures where multiple operators and vendors manage different interoperable NEs \cite{10329947}. This shift, supported by groups like Open Radio Access Network (O-RAN) and the Telecom Infra Project (TIP), aims to mitigate vendor lock-in and foster competition and innovation.
Data sharing between different vendors or operators is often restricted in disaggregated networks due to confidentiality concerns which create a barrier to centralized ML training.

\begin{figure}[h!]
	\centering
	\includegraphics[width=0.80\columnwidth]{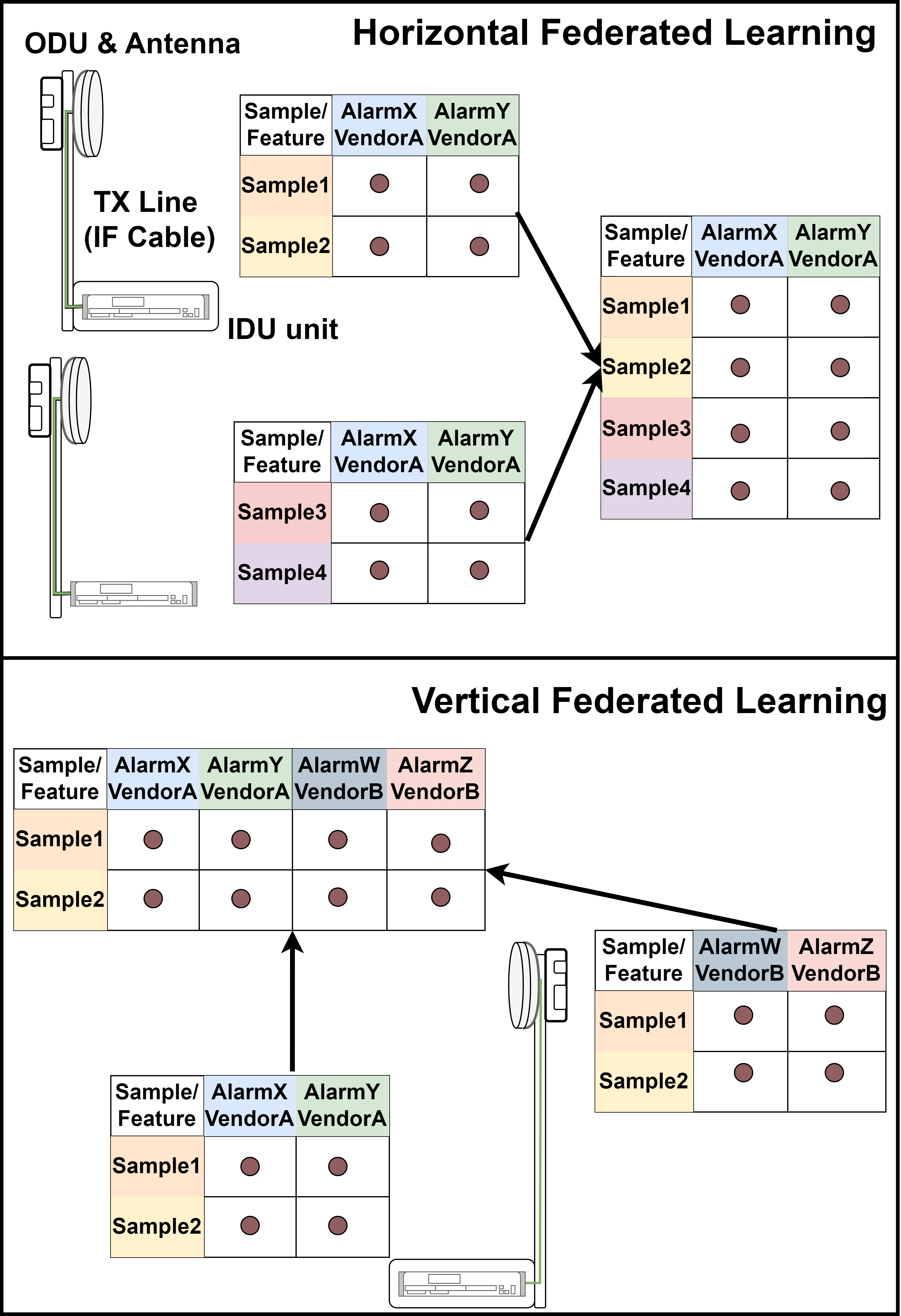}
	\caption{HFL in Single-Vendor HFL Scenario and VFL Multi-Vendor Scenario}
	\label{fig:HFLvsVFL}
        \vspace{-20pt}
\end{figure}

Federated Learning (FL) is a collaborative ML technique, in which each participating client, such as a vendor or an operator, trains a local model and shares only model updates with a central server that aggregates the local models as a third-party collaborator and then produces a global model that is again shared with local participating clients. By training models locally and simultaneously, and by only sharing model updates rather than raw data, the exposure of risk-sensitive data is minimized and a certain degree of data privacy can be guaranteed. The two main FL categories, based on how data is partitioned across clients, are Horizontal Federated Learning (HFL) and Vertical Federated Learning (VFL), where data is partitioned sample-wise and feature-wise, respectively. 
These categories are illustrated in Fig. \ref{fig:HFLvsVFL}. HFL is appropriate in case features across clients are homogeneous \cite{9927592} such as in a single-vendor multi-operator scenario where each operator uses the same type of equipment. Conversely, VFL is appropriate for disaggregated networks where several vendors manage different interoperable Network Elements (NEs), e.g., Indoor Unit Hardware (HW-IDU), Outdoor Unit Hardware (HW-ODU), and Network Operating System Software (NOS-SW) in a microwave network.  Also, VFL is mostly applied in cross-silo scenarios \cite {yang2023surveyverticalfederatedlearning} where participants (e.g. entities such as banks, hospitals, vendors etc.) seek privacy protection without compromising performance. Among the other privacy-preserving algorithms (e.g., Secure-Multi Party Computation, Differential Privacy (DP), etc.), Homomorphic Encryption (HE) is a promising approach to be applied in conjunction with VFL due to its additive property that allows secure aggregation of client outputs \cite{yang2023surveyverticalfederatedlearning} without significant performance degradation and communication overhead.

To address concerns about sharing raw data, we propose to use Split Neural Networks (SplitNN) and Gradient Boosting Decision Tree (GBDT)-based VFL (FedTree) for the hardware failure-cause identification problem in disaggregated microwave networks, considering both a default and HE-assisted implementation, and we test the proposed approaches in a multi-vendor deployment scenario. 
To the best of our knowledge, this is the first time VFL is adopted to address failure-cause identification in disaggregated microwave networks. 
The main paper contributions are as follows: 
\begin{enumerate}
    \item We define the problem of VFL-based failure-cause identification in a multi-vendor disaggregated microwave network and make use of real microwave hardware-failure data \cite{datacentric}.
    \item  We propose to use GBDT and SplitNN VFL to perform failure-cause identification in multi-vendor scenarios and compare them to a benchmark comprising a Single Vendor Scenario (SVS), where failure-cause classification is performed in a centralized manner.
    \item We carried on experimental results that show how the proposed VFL solutions achieve a maximum average F1-Score of 96.44\%, closely matching the centralized result of 96.49\%.
\end{enumerate}


\section{Background and Related Works}
\subsection{Background on Disaggregated Microwave Networks}
\label{subsec:dissmic}
\begin{figure}[t]
	\centering
	\includegraphics[width=0.90\columnwidth]{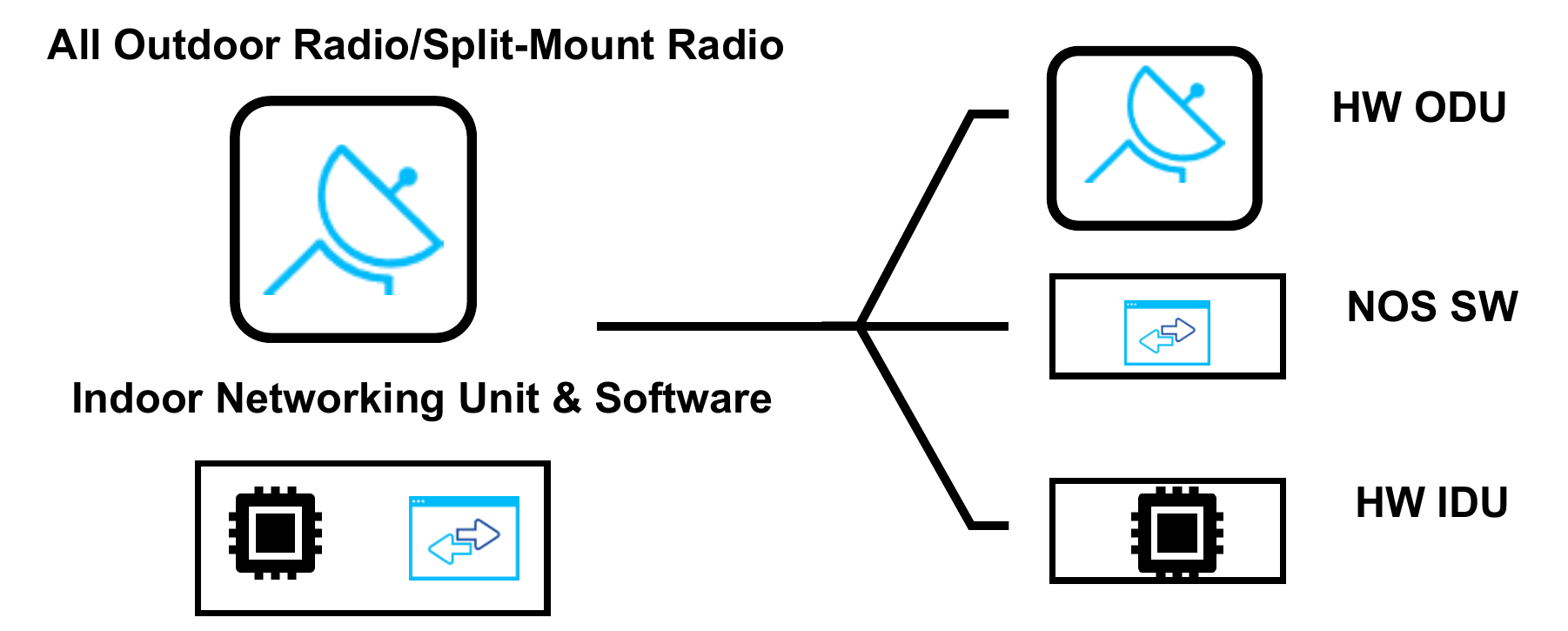}
	\caption{OSH's proposed disaggregation in microwave networks}
	\label{fig:microwavedisaggregated}
        \vspace{-20pt}
\end{figure}

The concept of disaggregation in communication networks has been investigated extensively in the last decade. Disaggregation mainly aims to reduce CapEx and foster innovation thanks to the definition of open interfaces that facilitate competition. The main research initiatives for disaggregation in wireless networks are the O-RAN Alliance and the TIP’s Wireless Backhaul Group (WBG), a project group of TIP that investigates disaggregation in wireless backhaul (i.e., microwave networks). WBG particularly works on disaggregated white box solutions for microwave networks and has recently presented a solution called OpenSoftHaul (OSH) \cite{TIP_Wireless_Backhaul}. As illustrated in Fig. \ref{fig:microwavedisaggregated}, OSH decouples microwave equipment into three distinct interoperable components, full outdoor radios in the microwave and millimetre-wave frequencies (i.e., HW ODU), IDU equipment that has open interfaces (i.e, HW IDU) and operator configurable Network Operating System (i.e., NOS SW).
Another significant initiative for disaggregation in the industry is Disaggregated Cell Site Gateways (DCSGs). A cell site gateway, or router, handles the aggregation of mobile data from cell sites and its transport to the service provider’s core network. Cell site routers (CSRs), also known as cell site gateways (CSGs), have traditionally been offered as integrated solutions by a limited number of large vendors, but this trend changes through disaggregation to avoid vendor lock-in. The main use cases for DCSG would be to perform 2G/3G/4G/5G mobile transport backhaul, but another DCSG  application is acting as IDU for Microwave Backhaul. In the case where IDU is disaggregated, a microwave modem is implemented as part of the ODU, reducing the IDU’s role to basic networking functions while ensuring seamless interoperability with other microwave equipment\cite{TelecomInfraProject2019}. 

\subsection{Related Works}
Recently, some ML-based solutions for predictive maintenance in microwave networks have been explored. For example, Ref.~\cite{9266116} illustrates a successful implementation (up to 93\% classification Accuracy) of supervised and unsupervised ML approaches for propagation failure identification (e.g., \emph{deep fading, interference}). Similarly, Ref.~\cite{datacentric, ASAP} shows successful implementation of ML-based classifiers (up to 96\% classification Accuracy) for hardware failure classification.   Ref.\cite{Ericsson2023Microwave} presents various preventive maintenance tasks, including hardware degradation detection, high-temperature early warning, and many specialized applications. 
Similarly, Ref.\cite{PAN2020106969} introduces the Proactive Microwave Anomaly Detection System (PMADS), a dynamic system designed for anomaly detection in cellular network microwave links. PMADS leverages network performance metrics and topological information to identify anomalies leading to faults. Other works (e.g., Refs.~\cite{AYOUB2022109466,9758095}) have investigated the application of eXplainable Artificial Intelligence (XAI) frameworks to provide insights into the decision-making process of ML classifiers. 

However, these works focus on monolithic scenarios, where data and models are managed by a single entity, ignoring disaggregation and collaborative approaches like FL. Few studies have explored FL-based solutions for microwave networks.
The most closely related work is Ref.~\cite{9927592}, and it shows the application of an HFL-assisted solution for identifying the causes of propagation failures (e.g., extra attenuation, deep fading) in microwave networks within a multi-operator non-disaggregated scenario. While HFL performs effectively, it requires that all clients in the federated setting share the same feature space. This is a strict limitation in disaggregated networks, where managed network elements may possess varying feature sets, as in our scenarios. 

While VFL-based solutions have been explored in other contexts, such as O-RAN~\cite{FTL_Ref2} and Network Function Virtualization (NFV) systems~\cite{10329604}, there is no prior work, to the best of our knowledge, that has investigated the application of VFL in disaggregated microwave networks.

\section{VFL-based Failure Cause Identification in Disaggregated Microwave Networks}

\subsection{Problem Definition}
The basic components of a microwave link are illustrated in Fig. \ref{fig:HFLvsVFL}. In this study, we consider split-mount configuration, where electronics are split into outdoor and indoor units of microwave radio (ODU and IDU, respectively). 
We model the hardware failure cause identification as a multi-class ML problem, where each class corresponds to a specific hardware failure (see next subsection). The classifier is trained collaboratively by vendors on partitioned features, as described in Section \ref{subsec:alarmgrouping}. In brief, each client has its specific dataset of features, i.e., IDU, ODU, NOS group features or combined versions, and the model predicts one of the four output classes that correspond to failure causes.

\subsection{Dataset Description}
\begin{figure}[t!]
	\centering
	\includegraphics[width=0.90\columnwidth]{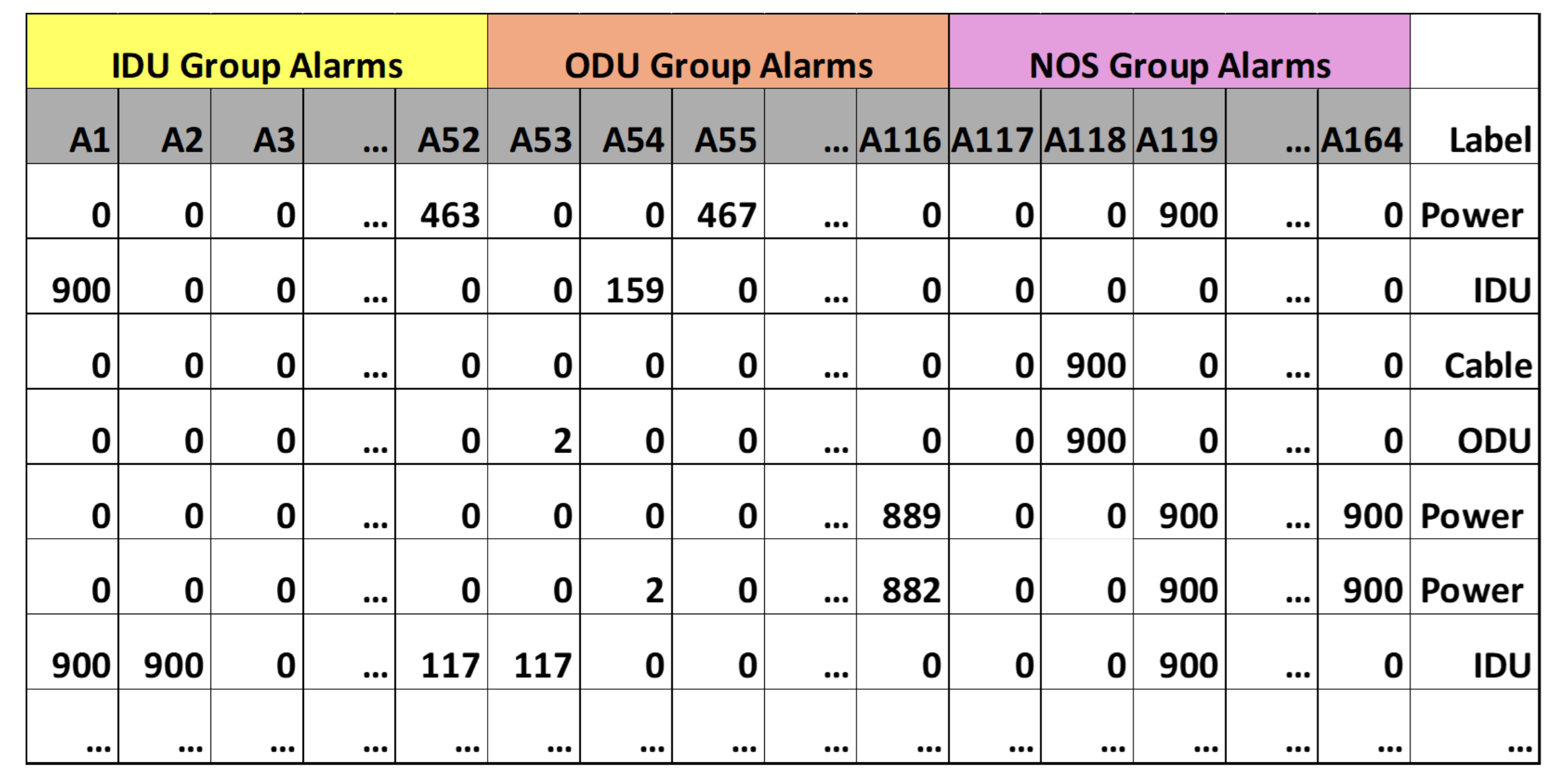}
	\caption{Illustration of the Dataset}
	\label{fig:microwavedataset}
    \vspace{-20pt}
\end{figure}

The NMS records equipment alarms for each microwave link. Our dataset was collected on a real microwave network by SIAE Microelettronica, and it is an open-source dataset 
\cite{datacentric} comprising 164 equipment alarm features ($A_1$ to $A_{164}$) and 1669 data points (observations). Each observation in the dataset compiles data from 15-minute (900 seconds) non-overlapping intervals. For each microwave link and for each 15-minute interval, the classification problem is fed with all features, each representing the number of seconds a specific alarm was active within that window. 

These features range from 0 to 900, where 0 means the alarm was OFF and 900 means the alarm was ON for the entire 15-minute period. Each observation is labelled and validated by domain experts. The dataset includes four types of hardware failures (i.e., labels) and the distribution of each class is as follows:
\begin{itemize}
    \item \textbf{Class-1 (IDU failure):} 515 observations (e.g., failure of an electronic IDU component, or a temperature issue due to improper equipment installation and/or a worn fan)
    \item \textbf{Class-2 (ODU failure):} 611 observations (failure causes are similar to those seen in the IDU in the point above)
    \item \textbf{Class-3 (Cable failure):} 207 observations (e.g., damaged connectors)
    \item \textbf{Class-4 (Power failure):} 336 observations (e.g., due to a power outage and/or a battery problem)
\end{itemize}

\subsection{Preprocessing}
After removing data for alarms with all zeros in the dataset (as they do not provide information), a total of 119 features are considered. Alarms can be considered as {a) binary, i.e., 1 if the alarm is ON and 0 if the alarm is OFF,} {b) categorical, i.e., 0 if the alarm is OFF, 1 if the alarm is ON for less than 45 seconds in a 15-minute window, 2 if the alarm is ON for less than 450 seconds and more than 45 seconds in the 15-minute window, and 3 if the alarm is ON for more than 450 seconds in a 15-minute window} {c) raw, i.e., number ranging
between 0 and 900. Following the approach in previous work \cite{datacentric}, we consider categorical binning.}

\section{VFL Approaches for Multi-Vendor ML-Model Development}


VFL is a promising approach for developing ML models based on heterogeneous feature sets across different organizations. Most of the VFL algorithms are based on the Stochastic Gradient Descent (SGD) algorithm, which can leverage logistic regression, SVM, and NN models as potential candidates. NN-based architectures enhance data privacy thanks to the complex nonlinear functions these networks employ by default \cite{wei2024verticalfederatedlearningchallenges}. Additionally, tree-based models have also been investigated in the context of VFL in the last years\cite{cheng2021secureboostlosslessfederatedlearning}. Also, recent studies\cite{grinsztajn2022treebasedmodelsoutperformdeep,shwartzziv2021tabulardatadeeplearning} show that tree-based, especially GBDT-based, models are as powerful as NN-based models on small-sized (i.e. lower than 10k samples) datasets. In this study, we employ both NN-based and GBDT-based VFL approaches.

\begin{figure}[hbpt]
    \centering
    \includegraphics[width=1.05\columnwidth]{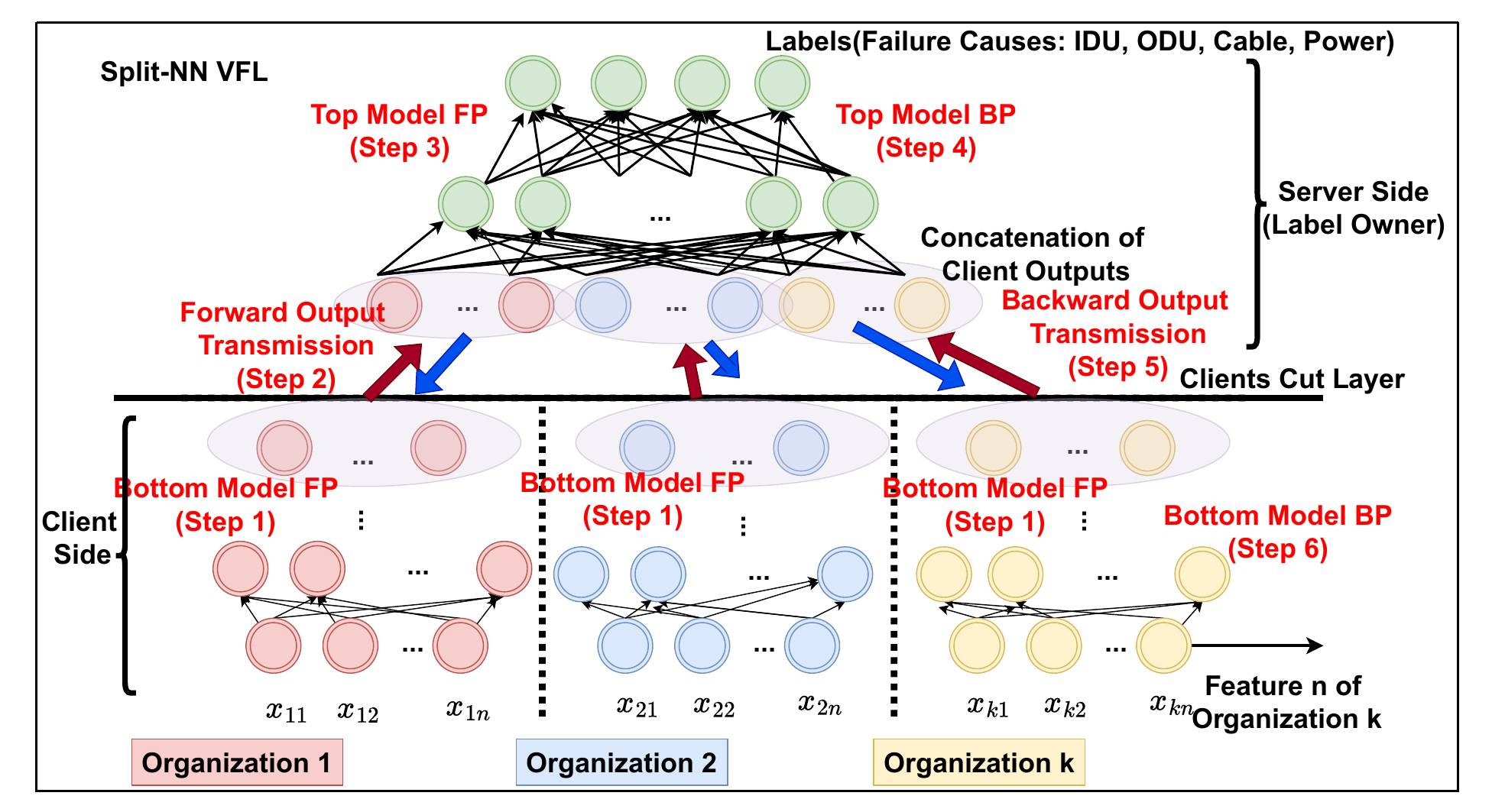}
    \caption{VFL Architecture Based on Split Neural Networks}
    \label{fig:SplitNN}
    \vspace{-15pt}
\end{figure}

\subsection{Split Neural Networks}
SplitNN is a VFL framework \cite{ceballos2020splitnndriven}, consisting of local NN models for \textit{k} clients (i.e., vendors) and a central server (i.e., a third-party entity). Each client manages its own local NN model, while the central server manages another NN model where the input layer is constructed by concatenating the local client's outputs. 
Fig.~\ref{fig:SplitNN}(a) shows the general architecture of SplitNN in the case of \emph{k} organizations, where each organization manages its local NN model. 
The first layer in each local model corresponds to the input layer and includes as many nodes as the number of features at each client; hidden layers follow, where the number of hidden layers and neurons per layer may differ among local models; the last layer is the output of the local NN models. The outputs of the local NN models are then concatenated and serve as inputs to the NN model at the central server, followed by \emph{h} hidden layers, and the last layer constituted by \emph{o} neurons serving as output. In our scenario, we consider \emph{o}=4, as the number of failure causes to discriminate.
The main steps of the SplitNN VFL algorithm can be summarized as follows \cite{wei2024verticalfederatedlearningchallenges}:

\begin{itemize}
    \item \textbf{Step 1: Bottom Model Forward Propagation (FP)}. Each client computes the FP of their local model using local data to obtain intermediate representations (\emph{Bottom Model FP} in Fig.~\ref{fig:SplitNN}(a)). This step is similar to conventional training except for loss calculation since labels are held in the global model. 
    \item \textbf{Step 2: Forward Output Transmission}. The outputs (intermediate cuts) from the local models (client side) are transmitted to the global model (server side). Advanced security protocols such as HE could be implemented at this step to provide privacy preservation.
    \item \textbf{Step 3: Top Model FP}. The server concatenates the received forwarded outputs and processes them through the global model for the final output of the classification task. There are various merging techniques available, including element-wise max, min, average pooling, product, and sum. While concatenation is the simplest approach, our experiments showed these techniques generally result in only slight differences in performance.
    \item \textbf{Step 4: Top Model Backward Propagation (BP)}. Based on the calculated loss from the prediction and the labels, the top model begins the backward propagation to compute gradients. 
    \item \textbf{Step 5: Backward Output Transmission}. Gradients computed in the top model are sent back to each participating client. 
    \item \textbf{Step 6: Bottom Model BP}. Upon receiving the gradients, each client updates model weights on their local model through backward propagation. This completes the cycle of one training iteration, referred to as \emph{epoch}.
\end{itemize}

\subsection{FedTree} 
FedTree is an alternative implementation of GBDT-based VFL algorithm \cite{MLSYS2023_3430e705}, where histograms (i.e., summation of gradients) are used as exchanged information differently from other implementations.
\begin{algorithm}[h]
\caption{Vertical FedTree (adapted from \cite{MLSYS2023_3430e705})}
\label{alg:VerticalFedTree}
\textbf{Input:} Clients $P_1, \ldots, P_n$, number of trees $T$, tree depth $d$. Assume $P_1$ has the labels.\\
\textbf{Output:}The final GDBT model f.\\
\small 
\begin{algorithmic}[1]
\STATE $S \leftarrow \text{ProposeSplitCandidates()}$ 
\STATE $P_1$ generates HE key pair $(K_{\text{pub}}, K_{\text{pri}})$.
\STATE $P_1$ sends public keys to the other clients.
\FOR{$i = 1$ to $T$}
    \STATE \text{Conduct on client $P_1$}
    \STATE $g, h \leftarrow \text{UpdateGradients}()$
    \STATE $[g], [h] \leftarrow \text{Encrypt}(g, K_{\text{pub}}), (h, K_{\text{pub}})$
    \STATE Send $[g], [h]$ to the other client.
    \FOR{$j = 1$ to $d$}
        \FOR{$k = 1$ to $n$}
            \STATE \text{Conduct on each client $P_k$}
            \STATE $[G_k], [H_k] \leftarrow \text{ComputeHist}([g], [h], S_k)$
            \STATE Send $[G_k], [H_k]$ to $P_1$.
        \ENDFOR
        \STATE \text{Conduct on $P_1$}
        \STATE $G, H \leftarrow \text{Decrypt}([G_k], K_{\text{pri}}),([H_k], K_{\text{pri}})$
        \STATE $\mathbf{H} \gets \bigcup_{k=1}^{n} \mathbf{H}_k$
        \STATE Update depth $j$ of tree $f_i$ using $G, H$ 
    \ENDFOR
\ENDFOR
\end{algorithmic}
\end{algorithm}
Within each iteration, the GBDT model trains iteratively and constructs a new tree aiming at minimizing the residuals, which are the differences between the predicted outcomes and the actual target values. 
In Vertical FedTree (See Alg. \ref{alg:VerticalFedTree}), clients (i.e., vendors) have their feature-wise partitioned local datasets (i.e., IDU-group alarms, ODU-group alarms, NOS-group alarms) and the active client $P_1$ also has labels that act as a server. In Alg. \ref{alg:VerticalFedTree}, at first, the active client proposes cut points and generates public and private keys for HE (lines 1-3). The active client calculates gradients, since it has the labels (i.e., failure causes), and then sends them to the other parties in an encrypted way. For each depth $d$, each party (i.e., including $P_1$) computes local histograms and sends local histograms to the active party again in an encrypted way (lines 4-14). The active party receives local histograms and concatenates them after decrypting. Using aggregated histograms, the active client calculates the best-split points and lets all parties update the node. Iteration ends when $d$ reaches $max\_depth$ for each tree in each client. (lines 15-18)
In this study, we adopt HE to secure exchanged information such as gradients and histograms. Specifically, our HE implementation is additive, based on the Paillier cryptosystem, and considers the \emph{Honest but Curious} threat model \cite{MLSYS2023_3430e705}, where participants strictly follow protocol but may infer sensitive information about others (e.g. gradients and histograms). In the default setting (i.e., FedTree), we protect local data while in the HE-assisted one (i.e., FedTreeHE) both local data and exchanged information are protected assuming a trusted server. 

\section{Case Studies and Numerical Results}
\label{sec:caseStudiesAndResults}

\subsection{Alarm Grouping}
\label{subsec:alarmgrouping}
Considering disaggregation aspects discussed so far, alarms are grouped into IDU-Alarms (31 features), ODU-Alarms (54 features), and NOS-Alarms (34 features), mimicking a situation where each interoperable NE is managed by a different player in the disaggregated network, e.g., a different vendor for each NE. 
The alarms from the electronic parts of the IDU are included in IDU-Alarms (e.g., modulator/demodulator), ODU-Alarms include radio link-related alarms as well as electronic components (e.g., voltage-controlled oscillator), whereas alarms that indicate SW-configuration mismatch and system-health like alarms are grouped into NOS-Alarms.

\begin{figure}[hbpt]
    \centering
    \vspace{-10pt}
    \includegraphics[width=1.0\columnwidth]{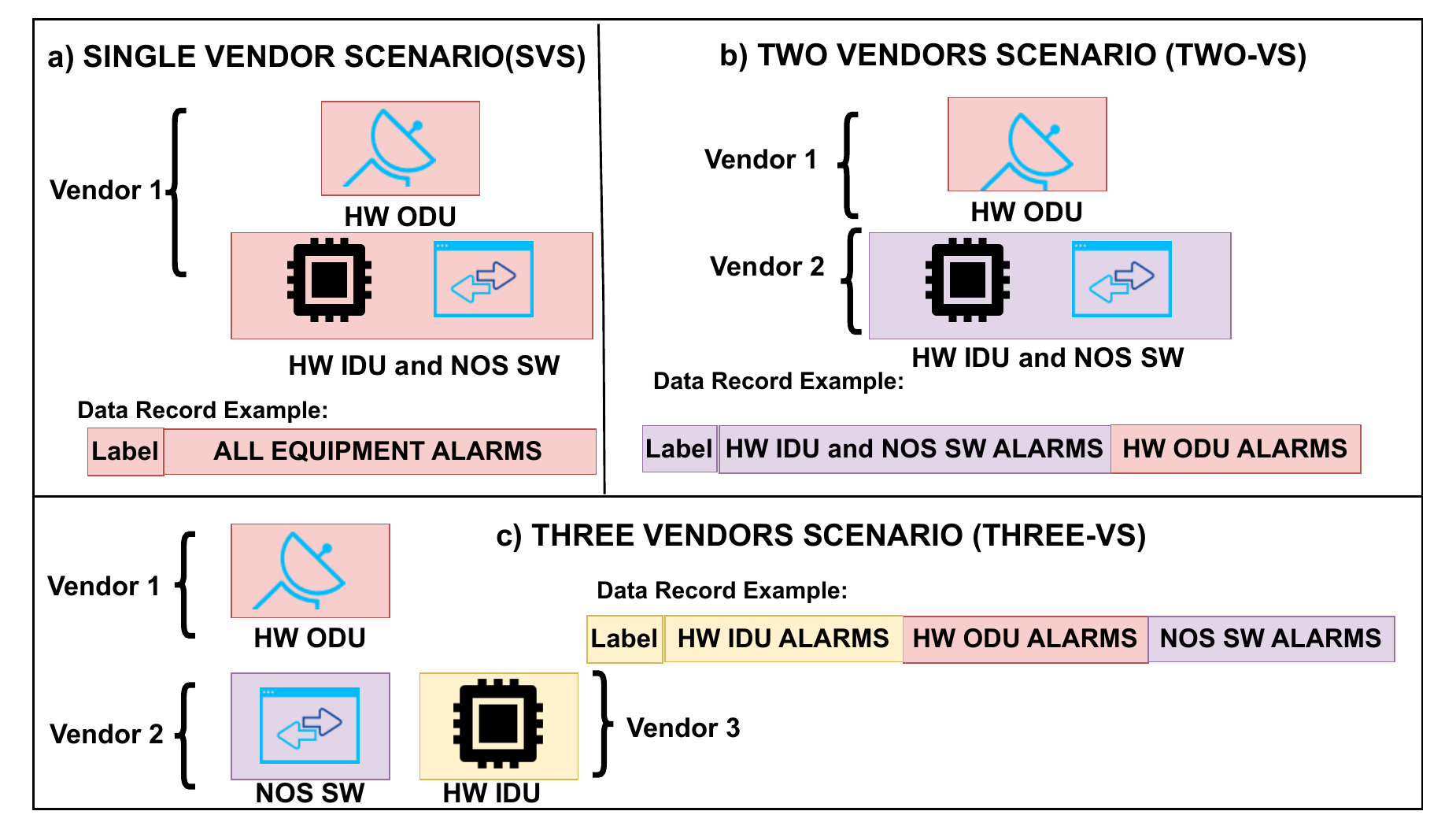}
    \vspace{-10pt}
    \caption{Single and Multi-Vendor Scenarios in Microwave Networks}
    \label{fig:DeploymentScenarios}
    \vspace{-15pt}
\end{figure}

\subsection{Deployment Scenarios}

We consider three different deployment scenarios, as shown in Fig. \ref{fig:DeploymentScenarios}: 1) Single Vendor Scenario (SVS), 2) Two Vendor Scenario (2-VS), and 3) Three Vendor Scenario (3-VS). 
\begin{itemize}
    \item \textbf{SVS}: A single vendor manages the whole network elements. All equipment alarm features (e.g., 119 alarm features) belong to one vendor.
    \item \textbf{2-VS}: HW-IDU and NOS-SW belong to one vendor, and HW-ODU belongs to another vendor. Vendor 1 has only alarm information related to ODU equipment (e.g., radio power setting alarms), while Vendor 2 has both HW-IDU and NOS-SW alarms (e.g., IDU temperature, software unit mismatch alarms).
    \item \textbf{3-VS}: Equipment is fully disaggregated, and each interoperable NE is managed by a vendor. 3-VS complies with the OpenSoftHaul architecture, where HW-IDU, HW-ODU, and NOS-SW are managed by different vendors.
\end{itemize}

\textbf{Baseline and FL methodologies}:
We consider \textit{six baselines: two centralized and four non-cooperative} and \textit{six FL scenarios}, obtained by varying the model type (i.e., NN-based or GBDT-based), the single vs. multi-vendor deployment (i.e., SVS, 2-VS, 3-VS), the privacy-preserving method (i.e., using HE or not), and whether vendors are cooperative or not.

Briefly, FL settings focus on collaboratively developing ML models across multiple vendors, allowing them to work together without sharing sensitive data. The centralized scenario (i.e., SVS) involves a single vendor developing the ML model using centralized data. In contrast, non-cooperative scenarios (i.e., 2VS and 3VS that use XGB or NN for local models)indicate that each data owner (e.g., IDU vendor) independently develops local ML models using only their data (e.g., 31 features for IDU vendor) without cooperation. 

\subsection{Evaluation Settings}
We perform stratified 5-fold cross-validation on both baselines and FL settings to find the best-performing model and report the mean Accuracy and F1-score, and standard deviation, over the 5 folds. Moreover, we consider hyperparameter optimization Grid Search (GS), for a fixed number of combinations, and select the best based on Accuracy. 
Additionally, for non-cooperative settings, we trained and found the best local models for each owner in the same way and averaged the best results over different owners (i.e., IDU, ODU, NOS vendors) and standard deviation capturing variation across these different owners' local models.

\subsection{Results Discussion}
\begin{figure}[t]
	\centering
	\includegraphics[width=0.95\columnwidth]{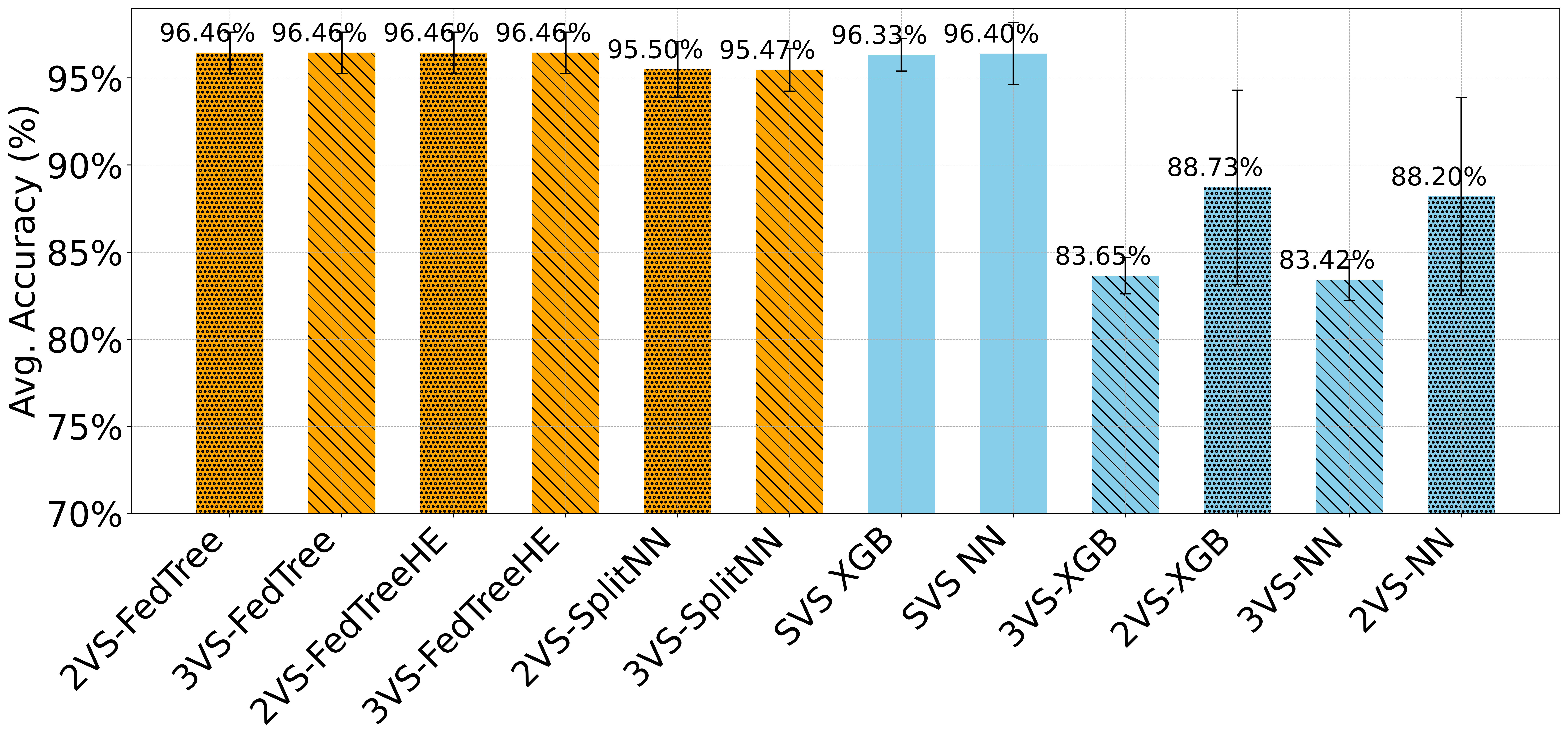}
    \vspace{-15pt}
	\caption{Avg. Accuracy in \% for VFL scenarios and baselines}
	\label{fig:microwaveScenarioAccuracy}
    \vspace{-10pt}
\end{figure}

\begin{figure}[t]
	\centering
	\includegraphics[width=0.95\columnwidth]{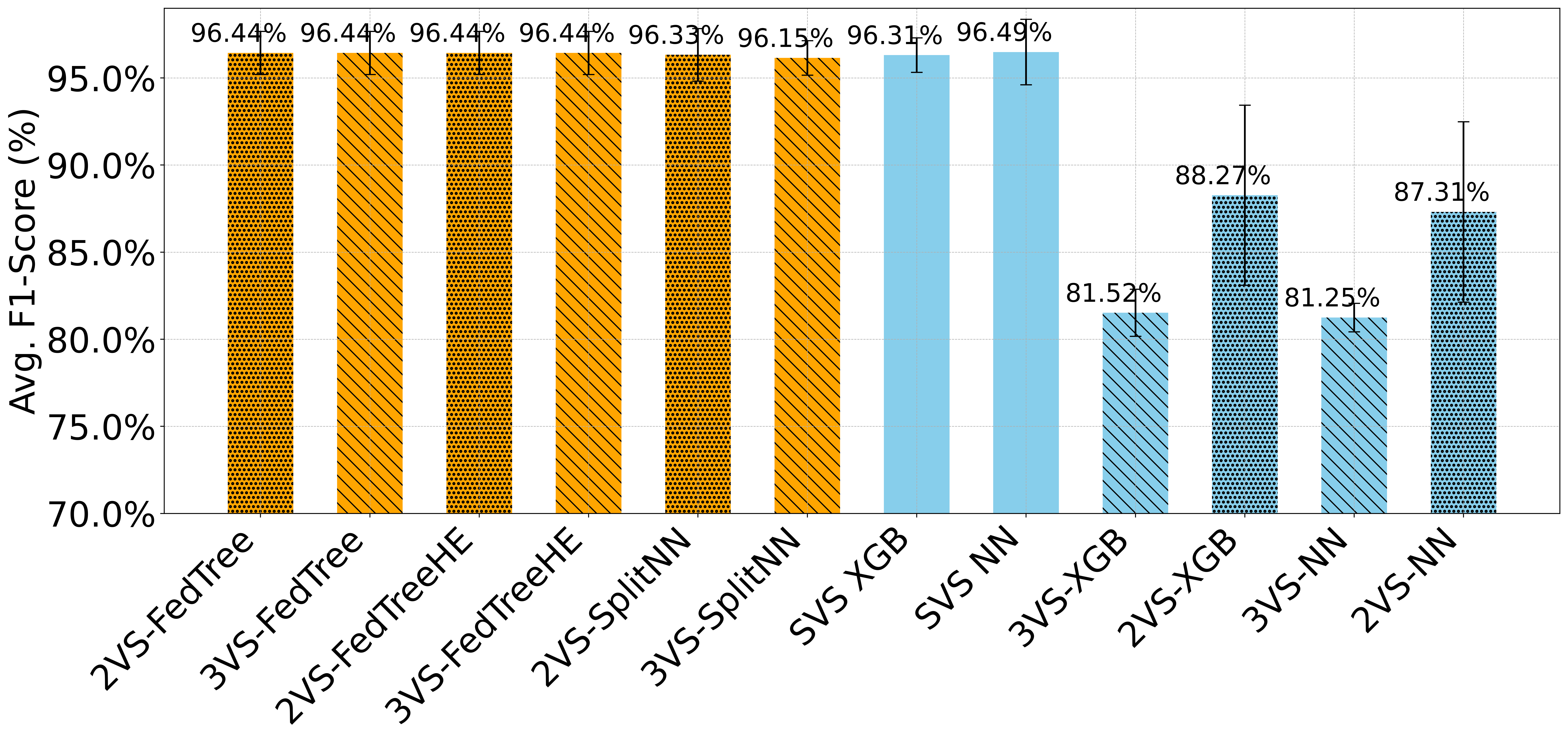}
    \vspace{-15pt}
	\caption{Avg. F1-Score in \% for VFL scenarios and baselines}
	\label{fig:MicrowaveScenarioF1Score}
    \vspace{-20pt}
\end{figure}

Figures \ref{fig:microwaveScenarioAccuracy} and \ref{fig:MicrowaveScenarioF1Score} show the Accuracy and F1-score (with Std Dev) for all proposed VFL and baseline scenarios. For the non-cooperative scenario, the results reflect the mean performance of the best local models from each data owner. Specifically, the 3-VS-XGB metrics represent the mean performance across individual models from IDU, ODU, and NOS, with the standard deviation capturing variation across these different owners. Due to space constraints, per-class metrics are not detailed, but class imbalance on the dataset causes no significant degradation, with performance variation across classes under 5\% in all cases.
Note that XGB and FedTree are addressed together as GBDT-based when discussing results. 

\noindent \textbf{Multi-Vendor vs. Single Vendor}. We observe that both 2-VS and 3-VS FL scenarios reach up to 96.46\% in Accuracy and 96.44\% in F1-Score and SVS scenarios reach up to 96.40\% in Accuracy and 96.49\% in F1-Score. SVS-NN and SVS-XGB centralized models perform similarly on their best-performing models (e.g., the difference is only 0.18\% in F1-Score). Both SplitNN and FedTree models show slight differences across all scenarios compared to centralized baselines, confirming VFL  effectiveness. Additionally, FL scenarios perform significantly better compared to non-cooperative scenarios (e.g., up to 13.04\% higher Accuracy and 15.19\% higher F1-Score for 3-VS, and up to 8.26\% higher Accuracy and up to 9.13\% higher F1-Score for 2-VS) where vendors only use their data to develop a local model without collaborating. 


\textbf{SplitNN vs. FedTree}.
GBDT-based models outperform NN-based models consistently both on Accuracy and F1-Score, except for the centralized scenario where SVS-NN performs slightly better than SVS-XGB with higher variance over folds. Moreover, FedTree model performance is not affected even if HE is integrated, which demonstrates that privacy can be guaranteed with no performance penalty.

\textbf{2-VS vs. 3-VS}.
Figures \ref{fig:microwaveScenarioAccuracy} and \ref{fig:MicrowaveScenarioF1Score} show a negligible performance difference between 2-VS-SplitNN and 3-VS-SplitNN, which demonstrates that the proposed approaches apply to two vendors and three vendors. In the case of non-cooperative, 2-VS performs better than 3-VS regardless of the model type used. This aligns with expectations since the IDU-NOS Vendor has a larger data chunk that drives performance metrics above (e.g., up to 5.31\% in Accuracy and 7.02\% in F1-Score).

\textbf{Training Time}.
Tab. \ref{tab:training_time} reports the training time in seconds for all considered scenarios. Depending on the model hyperparameters, VFL scenarios take longer training time compared to SVS (0.62\emph{s} vs 0.46\emph{s} in best case, and 77.41\emph{s} vs. 0.46\emph{s} in worst case). Among NN-Based scenarios, 2-VS takes the longest training time due to its more complex architecture (i.e., several hidden layers). Across the FedTree-based scenarios, 2-VS training time is longer compared to 3-VS. As a larger number of partitions results in smaller clients' data sizes, hence 3-VS has lower training time than 2-VS. Higher privacy protection increases training time, with FedTreeHE requiring up to 168 times more than its baseline, confirming that encryption increases complexity substantially. Encryption and decryption alone account for 15–30\% of total training time, even on smaller datasets.

\begin{table}[t]
\caption{Training Time (seconds)}
\vspace{-10pt}
\centering
\scriptsize
\setlength{\tabcolsep}{2pt} 
\begin{tabular}{|c|c|c|c|c|c|c|c|}
\hline
\multicolumn{3}{|c|}{\textbf{2VS}} & \multicolumn{3}{c|}{\textbf{3VS}} & \multicolumn{2}{c|}{\textbf{SVS}} \\
\hline
\textbf{SplitNN} & \textbf{FedTree} & \textbf{FedTreeHE} & \textbf{SplitNN} & \textbf{FedTree} & \textbf{FedTreeHE} & \textbf{NN} & \textbf{XGB} \\
\hline
9.05 & 0.62 & 64.84 & 3.53 & 0.46 & 77.41 & 1.70 & 0.46 \\
     &      & (14.82 HE) &      &      & (23.55 HE) &      &      \\
\hline
\end{tabular}
\label{tab:training_time}
\vspace{-20pt}
\end{table}
\vspace{-10pt}

\section{Conclusion}
We investigated the application of SplitNN-based and FedTree-based VFL for hardware failure-cause identification in disaggregated microwave networks. We consider equipment alarms from a real microwave network and compare multi-vendor (2-VS and 3-VS) to single-vendor deployment scenarios. The difference in Accuracy and F1-Score between centralized and federated scenarios is consistently within 1\%, regardless of the deployment strategies or model types used, proving that the proposed FL-based scenarios preserve classification performance while minimizing sensitive data leakage. In future work, a tailored privacy-preserving protocol will be developed to secure intermediate cuts in SplitNN. Additionally, DP will be applied to histogram concatenation on the server side in FedTreeHE scenarios to mitigate leakage risks, as the server has access to the global model and labels. These enhancements will strengthen the overall threat model.


\renewcommand\refname{References}
\addcontentsline{toc}{section}{References}
\bibliographystyle{IEEEtran}
\bibliography{ref}

\end{document}